\pdfoutput=1

\documentclass[prl,twocolumn,showpacs,preprintnumbers,amsmath,amssymb,
superscriptaddress,nofootinbib]{revtex4-1}
\usepackage{epsfig}
\usepackage{graphicx}
\usepackage{color}
\usepackage{hyperref}

\begin{document}

\title{Locality and Unitarity from Singularities and Gauge Invariance}

\author{Nima Arkani-Hamed}
\affiliation{School of Natural Sciences, Institute for Advanced Study, Princeton, NJ 08540, USA}

\author{Laurentiu Rodina}
\affiliation{Department of Physics, Princeton University, Jadwin Hall, Princeton, NJ 08540, USA}

\author{Jaroslav Trnka}
\affiliation{Center for Quantum Mathematics and Physics (QMAP), Department of Physics, University of California, Davis, CA 95616, USA}

\date{\today}

\begin{abstract}
We conjecture that the leading two-derivative tree-level amplitudes for gluons and gravitons can be derived from gauge invariance together with mild assumptions on their singularity structure. Assuming locality (that the singularities are associated with the poles of cubic graphs), we prove that gauge-invariance in just $(n-1)$ particles together with minimal power-counting uniquely fixes the amplitude. Unitarity in the form of factorization then follows from locality and gauge invariance. We also give evidence for a stronger conjecture: assuming only that singularities occur when the sum of a subset of external momenta go on-shell, we show in non-trivial examples that gauge-invariance and power-counting demand a graph structure for singularities. Thus both locality and unitarity emerge from singularities and gauge invariance. Similar statements hold for theories of Goldstone bosons like the non-linear sigma model and Dirac-Born-Infeld, by replacing the condition of gauge invariance with an appropriate degree of vanishing in soft limits.
\end{abstract}

\maketitle

\section{Gauge Redundancy}

The importance of gauge invariance  in our description of physics can hardly be overstated, but the fundamental status of ``gauge symmetry" has evolved considerably over the decades. While many older textbooks rhapsodize about the beauty of gauge symmetry, and wax eloquent on how ``it fully determines interactions from symmetry principles", from a modern point of view gauge invariance can also be thought of as by itself an empty statement. Indeed {\it any} theory can be made gauge-invariant by the ``Stuckelberg trick"--elevating gauge-transformation parameters to fields--with the ``special" gauge invariant theories distinguished only by realizing the gauge symmetry with the fewest number of degrees of freedom.

Instead of gauge symmetry we speak of gauge ``redundancy" as a convenient but not necessarily fundamental way of describing the local physics of Yang-Mills and gravity theories.  Indeed in the sophisticated setting of quantum field theories and string theories at strong coupling we have seen the crucial importance of understanding gauge symmetries as ``redundancies"--for instance, in the famous gauge-gravity duality, it is silly to ask ``where is the gauge symmetry?" in the bulk or ``where is general covariance" on the boundary; these are merely two differently-redundant descriptions of the same physical system.

If gauge ``symmetries" are merely redundancies, why have they been so useful? We can see the utility of gauge-redundancy  \cite{WeinbergQFTvol1} in the down-to-earth setting of scattering processes for elementary particles even at weak coupling, where we encounter a peculiarity in the Poincare transformation properties of scattering amplitudes. When the momenta of particles are transformed, the amplitude transforms according to the {\it little group}. Thus e.g. in four dimensions under a Lorentz transformation $\Lambda$ the amplitude picks up phases $e^{i h \theta(\Lambda,p)}$ for each massless leg of momentum $p$ helicity $h$, and an $SO(3)$ rotation on the massive particles. On the other hand, the standard formalism of field theory, the amplitudes are computed using Feynman diagrams, which give us ``Feynman amplitudes" that are not the real amplitudes, but are instead Lorentz tensors.  We contract them with {\it polarization vectors} to get the actual amplitudes--the polarization vectors are supposed to transform as ``bi-fundamentals" under the Lorentz and little groups. For  massive particles of any spin, there is a canonical way of associating polarization vectors with given spin states. But this is impossible for massless particles. Say for massless spin 1, we associate $\epsilon_\mu^{\pm}(p)$ with the $\pm$ polarizations of photons: the $\epsilon_\mu$ do not transform as vectors under the Lorentz group. Indeed consider Lorentz transformations $\Lambda$ that map $p$ into itself $(\Lambda p)_\mu = p_\mu$. Then, it is trivial to see that $(\Lambda \epsilon)$ does not equal $\epsilon$ in general, rather we find $(\Lambda \epsilon)_\mu = \epsilon_\mu + \alpha(p) p_\mu$. Thus the polarization vector itself does not transform properly as a four-vector,  only the full equivalence class $\{\epsilon_\mu| \epsilon_\mu \sim \epsilon_\mu + \alpha(p) p_\mu\}$ is invariant.  These are all the ``gauge-equivalent" polarization vectors. And so, for the amplitude obtained by contracting with $\epsilon$'s to be Lorentz-invariant, we must have that under replacing $\epsilon_\mu \to p_\mu$ the amplitude vanishes; i.e. we must satisfy the ``on-shell Ward identity' $p^\mu M_{\mu ...} = 0$. In order to guarantee that the Lorentz tensors $M_{\mu_1 \cdots \mu_n}$ arising from Feynman diagrams from a Lagrangian satisfies this on-shell Ward-identity,  the Lagrangian must be carefully chosen to have an (often non-linearly completed) gauge-invariance, which is then gauge-fixed. From the  modern point of view, then, gauge symmetry is merely a useful redundancy for describing the physics of interacting massless particle of spin 1 or 2, tied to the specific formalism of Feynman diagrams, that makes locality and unitarity as manifest as possible.

But over the past few decades, we have seen entirely different formalisms for computing scattering amplitudes not tied to this formalism, and here gauge redundancy makes no appearance whatsoever. Instead of polarization vectors that only redundantly describe massless particle states, we can use spinor-helicity variables $\lambda_a$, $\widetilde{\lambda}_a$ for the $a$'th particle, with momentum $p_{a}^{\alpha \dot{\alpha}} = \lambda_a^\alpha \tilde{\lambda}^{\dot{\alpha}}$. The $\lambda, \tilde \lambda$'s {\it do}  transform cleanly as bi-fundamentals under the Lorentz and little groups; under a Lorentz transformation $\Lambda$ that maps $(\Lambda p) = p$, we have $\lambda \to t \lambda, \tilde{\lambda} \to t^{-1} \tilde{\lambda}$. Thus while the description of amplitude using polarization vectors is gauge-redundant, the amplitude is directly a function of spinor-helicity variables, with the helicities encoded in behavior under rescaling $M(t_a \lambda_a, t_a^{-1} \tilde{\lambda}_a) = t_a^{-2 h_a} M(\lambda_a,\tilde{\lambda}_a)$.

With this invariant description of the fundamental symmetries and kinematics of amplitudes at hand, it becomes possible to pursue entirely new strategies for determining the amplitudes. In a first stage, one can speak of a modern incarnation of the S-matrix program, where the fundamental physics of locality and unitarity are imposed to determine the amplitudes from first principles. This has allowed the computation of amplitudes in an enormous range of theories, from Yang-Mills and gravity to goldstone bosons, revealing stunning simplicity and deep new mathematical structures that are completely hidden in the usual, gauge-redundant Feynman diagram formalism. Conversely and more ambitiously, these developments suggest that what we think of as ``scattering amplitudes from local evolution in spacetime" might fundamentally be something entirely different: instead of merely {\it exploiting} locality and unitarity to determine the amplitudes, we seek ``scattering amplitudes" as the answer to very different natural mathematical questions, and only later {\it discover} that the results are local and unitary. Carrying this program out in full generality for all interesting theories would likely shed powerful new light on a deeper origin for both space-time and quantum mechanics itself.

A step in this direction has been taken with the discovery of the ``Amplituhedron" \cite{Arkani-Hamed:2013jha}, a geometric object generalizing plane polygons to higher-dimensional spaces, whose ``volume" computes scattering amplitudes for maximally supersymmetric four-dimensional theories in the planar limit (in particular giving tree-level gluon scattering amplitudes for the real theory of strong interactions relevant for particle collisions at the LHC). In this example we can see concretely how the usual rules of spacetime and quantum mechanics emerge from more primitive principles. 

\section{Role Reversal}

In this letter, we will explore aspects of locality and unitarity from a point of view entirely orthogonal to these recent developments. As emphasized above, much of the explosion of progress in understanding scattering amplitudes has taken place precisely by eschewing any reference to gauge-redundancy, and working directly with the physical on-shell amplitudes. Here we instead return to the requirement of on-shell gauge invariance as primary, and consider rational functions built out of polarization vectors and momenta, without making any reference to an underlying Lagrangian, Feynman rules or diagrams  of any kind. Surprisingly, we find that with  mild restrictions on the form of functions we consider, the requirement of on-shell gauge-invariance alone uniquely fixes the functions to match the tree amplitudes of Yang-Mills theory for spin one and gravity for spin two. There is a similar story determining the amplitudes for goldstone bosons of the non-linear sigma model and the Dirac-Born-Infeld action, where the requirement of on-shell gauge invariance is replaced by an appropriate vanishing of amplitudes in soft-limits.

Suppose that we are handed a rational function of momenta and polarization vectors. What constraints determine this function to correspond to ``scattering amplitudes"? One might imagine that both locality and unitarity are crucially needed for this purpose. In other words, we have to assume that this function has only {\it simple} poles when the sum of a subset ${\cal S}$ of the momenta $P^\mu_{\cal S}=\sum_{i \subset {\cal S}} p^\mu_i$ goes on-shell i.e. the only singularities look like $\sim 1/P_{\cal S}^2$, {\it and} that the function factorizes on the poles into the product of lower-point objects on the left and right, with an extra intermediate line. Note that locality and unitarity are intertwined in an interesting way. Factorization on {\it simple} poles guarantees that (in Lorentzian signature with the Feynman $i \epsilon$'s included) the imaginary part of amplitudes correspond to particle production. But factorization also implies that the singularities must be associated with a graph structure: sitting on a factorization channel, we can seek further singularities to deeper channels, but the longest sequence of poles we can encounter in this way all correspond to the $(n-3)$ propagators of some cubic graph.

The expectation that both locality and unitarity are needed to fix the form of the amplitude comes from our direct familiarity with simple theories of scalars, like $\phi^3$ or $\phi^4$ theory. If we only impose poles when $P^2 \to 0$ and the usual mass dimensions of amplitudes associated with, say, $\phi^3$ theory, nothing forbids the presence of various trivially ``illegal" terms of the form e.g. for $n=5$
\begin{equation}
\frac{1}{(p_1 + p_2)^2 (p_2 + p_3)^2}, \frac{1}{((p_1 + p_2)^2)^2}
\end{equation}
The first term has legal simple poles, but in overlapping channels in a way that never arises from Feynman diagrams; thus while at the coarsest level it's singularities are ``local poles" it doesn't correspond to any local spacetime process. The second doesn't suffer from overlapping poles but has double poles. We can choose to also enforce locality by declaring that our functions can only have the poles corresponding to cubic graphs. If we again imagine objects with the mass dimension corresponding to a $\phi^3$ theory, we would get a sum over cubic graphs $\Gamma$ with some numerical coefficient $n_\Gamma$:
\begin{equation}
\sum_{\Gamma} \frac{n_\Gamma}{D_\Gamma}
\end{equation}
where $D_\Gamma$ is the product of the propagators of the cubic graph $\Gamma$. This expression corresponds to the amplitude only if the coefficients $n_\Gamma$ are all equal but this is obviously not an automatic consequence of our rules. We must demand unitarity--factorization into product of lower amplitudes--to force all the $n_\Gamma$ to be equal.

Our central claim in this note is that while locality and unitarity must be imposed to determine amplitudes for garden-variety scalar theories, {\it much} less than this is needed to uniquely fix the function to be ``the amplitude" for gauge theories and gravity. In fact, we conjecture that simply specifying that the only singularities occur when the sum of a subset of momenta goes on-shell $P^2 \to 0$, together usual power-counting (which also enforces {\it non-trivial} gauge invariance)  uniquely fixes the function!
We will sketch the essential ideas in this note, a more detailed exposition of our proof and other related results will appear in \cite{BCFWprep}. Other observations about the surprisingly restrictive power of on-shell gauge invariance have recently been made in
\cite{Boels:2016xhc}.

To begin with, we can enforce only locality, in the form of the location of singularities of the amplitudes. This tells us to only look at functions whose singularities are (powers of) propagator poles appearing  cubic graphs, as we did above in the scalar case. But we don't demand unitarity: we don't ask the poles to be simple, and we don't demand that the function factorizes on the poles. We find that instead the leading non-trivial gauge-invariants with the singularities of cubic graphs are unique in both Yang-Mills and gravity, and give us the amplitude! The necessity of simple poles and factorization--and thus unitarity--thus follows from locality and gauge invariance. We will sketch a straightforward proof of this fact, which begins by showing that given the poles of cubic graphs, gauge-invariance alone (with no assumption about factorization) fixes the structure of the soft limit of any expressions to reproduce the usual Weinberg soft theorems \cite{Weinberg:1965nx}.

But we are making a stronger conjecture, that even the structure of singularities associated with cubic graphs need not be enforced: we need only assume that the singularities occur when $P^2_{\cal S} \to 0$. We will consider functions that have at most degree $\beta$ singularities of this form, that is our most general ansatz is
\begin{equation}
\sum_{\{{\cal S}_1,\cdots, {\cal S}_\beta\}} \frac{N^{(\alpha)}_i}{P^2_{{\cal S}_1}\cdots P^2_{{\cal S}_\beta}}
\end{equation}
Here $N(\alpha)_i$ is a polynomial in the momenta (and linear in all the polarization vectors), with a total of $\alpha$ momenta in the numerator.

We will now only ask for this expression to be on-shell gauge-invariant. Now clearly, even if there are no singularities at all i.e. $\beta = 0$, we can of course trivially build gauge-invariants simply starting with linearized field strengths $f_{\mu \nu} = p_\mu \epsilon_\nu - p_\nu \epsilon_\mu$,and contracting $n$ of these together in any way we like. This would give us a number $\alpha \geq n$ of momenta. These correspond to the amplitudes from local higher-dimension operators. We will thus ask that our functions are {\it non-trivially} gauge invariant, and so we will demand that $\alpha < n$; the only hope for making gauge-invariants now crucially must crucially use momentum-conservation $p_1^\mu + \cdots p_n^\mu=0$. It is then easy to see that this is impossible with purely local expressions, and we must allow poles so $\beta > 0$. Our precise claim is that it is impossible to build a gauge-invariant unless $\alpha = (n-2)$ for gauge-theory and $\alpha = 2 (n-2)$ for gravity, and furthermore this is impossible for all $\beta=0,1,\cdots,(n-4)$, but that there is a unique gauge-invariant at $\beta = (n-3)$. In fact just demanding gauge-invariance in $(n-1)$ legs suffices to fix the function. This unique object picks out the singularities from cubic graphs and factorizes on poles; locality and unitarity arise from singularities and gauge invariance. While we haven't yet completed a proof of this conjecture, we will show how it works in some non-trivial examples which suggest the structure a proof should take.

All of these statements are made in general $D$ spacetime dimensions: we are simply working with lorentz-invariants multilinear in the polarization vectors, of the form $(\epsilon_i \cdot \epsilon_j)$, $\epsilon_i \cdot p_j$ and $(p_i \cdot p_j)$, only satisfying the relations $p_i^2 = 0, \epsilon_i \cdot p_i=0$ and momentum-conservation $\sum_i p_i^\mu = 0$. Thus the gauge-invariance checks where we demand the vanishing of the amplitude upon substituting $\epsilon_i^\mu \to p_i^\mu$ can only follow from these relations.

\section{Locality, Unitarity and Gauge Invariance}

Let us begin by focusing on the tree-level scattering amplitudes in Yang-Mills theory; we will later summarize the precisely analogous statements for gravity. The group structure of gluon amplitudes can be stripped off in trace factors ${\cal A}_n = \sum_{\sigma/Z} {\rm Tr}(T^{\sigma_1}T^{\sigma_2}\dots T^{\sigma_n})A_n(123\dots n)$, where $A_n$ is an ordered amplitude which is a gauge invariant cyclic object.  All poles in $A_n$ are local cyclic factors, $P_{ij}^2=(p_i+p_{i+1}+\dots p_j)^2$.  And on these poles $A_n$ factorizes as a product of two ordered amplitudes,
\begin{equation}
\lim_{P^2\rightarrow0} A_n  = \sum_{h} A_L^{(h_L)}\frac{1}{P^2} A_R^{(h_R)}\label{unit}
\end{equation}
were we sum over all internal degrees of freedom $h$. In practice we can replace the helicity sum over the intermediate line $I$ by $\sum_h \epsilon_{I, h}^{\mu} \epsilon_{I, -h}^\nu \to \eta^{\mu \nu}$; this differs from the true polarization sum by terms proportional to $p_I^\mu, p_I^\nu$ which vanish by gauge invariance, when contracted into the lower-point  amplitude factors.

The cyclic amplitude $A_n$ can by calculated using color-ordered Feynman rules. For each cubic graph $\Gamma$ we get
\begin{equation}
D_n^{(\Gamma)} = \frac{N_n^{(\Gamma)}(\epsilon_i,p_j)}{P_{\sigma_1}^2P_{\sigma_2}^2\dots P_{\sigma_{n-3}}^2}\label{FD}
\end{equation}
where all the factors $P_{\sigma_a}^2$ in the denominator come from Feynman propagators of cubic diagrams. The numerator is a polynomial in all polarization vectors $\epsilon_i$ and $n-2$ momenta $p_j$ and contains scalar products $(p_i\cdot p_j)$, $(p_i\cdot \epsilon_j)$ and $(\epsilon_i\cdot\epsilon_j)$. For the diagrams with four point vertices we get fewer than $n-3$ propagators but they can be also put (non-uniquely) in the form by multiplying both numerator and denominator by some $P^2$.

Feynman diagrams are designed to make locality and unitarity as manifest as possible, but gauge-invariance is not manifest diagram-by-diagram: we have to sum over all Feynman diagrams to get a gauge invariant expression. The tension between locality, unitarity and gauge invariance is vividly seen in the four-particle amplitude. The color ordered amplitude $A_4$ is a sum of three Feynman diagrams, schematically written as (ignoring all indices)
\begin{equation}
A_4 \sim \frac{(\epsilon\cdot p)(\epsilon\cdot\epsilon)}{s}+\frac{(\epsilon\cdot p)(\epsilon\cdot\epsilon)}{t}+(\epsilon\cdot\epsilon)(\epsilon\cdot\epsilon)
\end{equation}
Only the sum of all three terms is gauge invariant which can be made manifest once we write $A_4$ as
\begin{equation}
A_4 \sim \frac{F^4}{st}
\end{equation}
where the numerator is just a (color-ordered) local amplitude. This expression is trivially gauge-invariant but we don't have {\it manifest} locality and unitarity: we see the {\it product} of $s t$ in the denominator. It is  impossible to write the amplitude as a sum over $s$ and $t$ channels in a way that is both Lorentz invariant and gauge invariant.

\section{Unitarity From Locality and Gauge Invariance}

Elaborating further on the example from the previous section we can ask what is the minimal number of momenta $p_j$ we need in order to make a polynomial in $\epsilon_1,\dots,\epsilon_n$ gauge invariant. Obviously, if we take $n$ momenta we can always build gauge invariant tensors $\epsilon^{[\mu}p^{\nu]} = (p_\mu\epsilon_\nu-p_\nu\epsilon_\mu)$ and contract $n$ of them in an arbitrary way. But can we can a non-trivial invariant, one which has fewer than $n$ momenta? This has a chance of being possible because of momentum conservation.  The first non-trivial case is with $n-2$ momenta $p_i$. It is easy to see that
if we demand the object is just a polynomial we find there exist no gauge invariant, but if we allow poles we certainly find at least one solution which is the amplitude $A_n$ written as a sum of Feynman diagrams.

Let us now consider a set of all cubic graphs with cyclic ordering of external legs and for each of them we write an expression $\widetilde{D}_n^{(\Gamma)}$ of the form (\ref{FD}) where the poles in the denominator are dictated by the internal lines of the given graph. Unlike in Feynman diagrams we do not demand the numerator comes from Feynman rules and therefore we are not imposing unitarity; more invariantly we are not asking the amplitude to actually factorize on factorization channels.  Instead we take $N_n^{(\Gamma)}$ to be an arbitrary polynomial of degree $n-2$ in momenta $p_j$ and $n$ polarization vectors $\epsilon_i$. For four point we get,
\begin{align}
N_4=& \alpha_1 (\epsilon_1\cdot p_2)(\epsilon_2\cdot p_3)(\epsilon_3\cdot\epsilon_4) + \alpha_2 (\epsilon_1\cdot p_2)(\epsilon_3\cdot p_4)(\epsilon_2\cdot\epsilon_4)\nonumber\\
&+\alpha_3 (p_1\cdot p_2)(\epsilon_1\cdot\epsilon_2)(\epsilon_3\cdot\epsilon_4) + \dots
\end{align}
We of course impose $(\epsilon_i\cdot p_i)=0$ and momentum conservation $\sum_i p_i = 0$.  The same structure of numerator is  used for the $s$ and $t$ channels, but with different parameters $\alpha_k^{(1)}$ and $\alpha_k^{(2)}$.  Now we now consider a sum of all expressions associated with graphs $\Gamma$,
\begin{equation}
\widetilde{A}_n = \sum_\Gamma \widetilde{D}_n^{(\Gamma)} \label{sum}
\end{equation}
and impose gauge invariance in $n-1$ legs. We claim that this specifies an unique expression which is an $n$ point tree-level amplitude, $\widetilde{A}_n = A_n$. Note we do not have to even check gauge invariance in the $n^{\rm th}$ leg, everything is fixed already.

The proof goes as follows: First, it is easy to show that if we do not consider momentum conservation then there are no non-trivial gauge invariants, we could have only many copies of $\epsilon^{[\mu}p^{\nu]}$. If we add momentum conservation there are more options. Let us now consider a polynomial with $k$ factors of  $\epsilon^{[\mu}p^{\nu]}$. This is trivially gauge invariant in $k$ legs. Using momentum conservation we can gain one ``free" gauge invariance, but only for $k\geq n-2$ for scalar function $B_n$ and $k\geq n-1$ for tensor function $B_n^{\mu\nu}$. Both statements can be proven quite easily by using momentum conservation and also classifying all tensor structures in $B_n^{\mu\nu}$.

We assume inductively that $\widetilde{A}_n=A_n$ is unique for $n$ particle case. Now we take the expansion (\ref{sum}) for $n+1$ particles and go to the soft limit of one of the particles, $p_{n+1}\equiv q\rightarrow0$. It is easy to show that gauge invariance requires the leading divergent term to be Weinberg soft factor,
\begin{equation}
\widetilde{A}_{n+1} = \left(\frac{\epsilon\cdot p_1}{q\cdot p_1} - \frac{\epsilon\cdot p_n}{q\cdot p_n}\right) B_n(p^{n-2}) + {\cal O}(1)
\end{equation}
where $B_n$ is the gauge invariant function in $n$ legs with $n-2$ powers of momenta which is $B_n=A_n$ by induction. The important point here is that the soft limit is controlled by the usual Weinberg soft factor {\it purely as a consequence of gauge invariance}, without any further assumption about factorization.

Now since both $A_{n+1}$ and $\widetilde{A}_{n+1}$ have equal leading pieces, we can consider instead the object $M_{n+1}=\widetilde{A}_{n+1}-A_{n+1}$, which has vanishing leading piece. This is important because a non-zero order $\mathcal{O}(z^m)$ has a contribution to order $\mathcal{O}(z^{m+1})$ through momentum conservation (see \cite{delta} for a discussion). Then the subleading piece in the soft limit has the form
\begin{equation}
\delta_1 M_{n+1} = \frac{\epsilon^\mu q^\nu B_n^{\mu\nu}(p^{n-2})}{q\cdot p_1} + \frac{\epsilon^\mu q^\nu \overline{B}_n^{\mu\nu}(p^{n-2})}{q\cdot p_n}
\end{equation}
where we omitted the terms with double poles which are directly ruled out by gauge invariance. The tensors $B_n^{\mu\nu}$, $\overline{B}_n^{\mu\nu}$ have $k=n-2$ and therefore are ruled out. At higher order terms in the soft limit we always get $\delta_p M_{n+1}\sim X_n^{\mu\nu}(p^{n-2})$ for some tensor $X$ which is then ruled out and all these terms must vanish.

It is interesting that in these arguments, it suffices to check gauge invariance only in $(n-1)$ legs to uniquely fix the answer! This observation explains why the object factorizes on poles. We'd like to determine what our unique gauge-invariant looks like on a factorization channel. Since there is already a unique gauge invariant only checking invariance on $(n-1)$ legs, we can take ``left" and ``right" gauge invariants ignoring gauge-invariance on the intermediate line; gluing together these unique objects then gives us something that is gauge-invariant in all $n$ legs, and therefore must match the unique $n$-pt gauge-invariant on this channel. This shows that gauge-invariants factorize on poles, allowing us to see the emergence of unitarity very directly.

\section{Locality from Gauge Invariance}

We showed that unitarity is a derived property of gluon amplitudes if we demand only locality and gauge invariance. But we can go even further and even remove the requirement of locality.
We again consider a sum of terms (\ref{sum}) but now we give up on th assumption that individual terms (\ref{FD}) have poles which correspond to cubic diagrams. We just consider any cyclic poles $P_{ij}^2=(p_i+p_{i+1}+\dots+p_j)^2$, and even allow powers $(P_{ij}^2)^{\#}$. The only assumption is that the total number of poles in the denominator (the degree of $P^2$) is $n-3$. For example, for $n=5$ case we allow terms of the form
\begin{equation}
\frac{N_5^{(1)}}{s_{12}^2},\quad \frac{N_5^{(2)}}{s_{12}s_{23}}\label{nonlocal}
\end{equation}
While the double (or higher) poles can come from the Lagrangians with non-canonical kinetic term the second term can not be associated with any local interaction as it does not correspond to any ``diagram'' of particle scattering. The numerator is an arbitrary polynomial in $n$ polarization vectors $\epsilon_i$ and $n-2$ momenta $p_j$.

We now conjecture that if we simply impose gauge invariance on the general sum of all possible terms with $n-3$ cyclic poles (\ref{sum}) the only solution is again only the $n$ point scattering amplitude $A_n$. There are no other solutions and all numerators for terms like (\ref{nonlocal}) are forced to vanish as a consequence of gauge invariance. We have directly checked this conjecture by brute force up to the n=5, which is already highly non-trivial. We will also give an analytic proof of the analog of this conjecture for the non-linear sigma model up to at $n=8$ points (which is the NLSM analog of $n=5$ for YM theory) below.

\section{Gravity and BCJ}

The story for gravitons is essentially identical. In particular, we can again consider cubic graphs
with no ordering of external legs. For each graph we associate an expression (\ref{FD}) to each of them. The denominator contains $n-3$ propagators consistent with the cubic graph -- the poles are not restricted to be cyclic sums of momenta anymore. The numerator $N_n^{(\Gamma)}$ is polynomial of degree $2(n-2)$ in momenta $p_i$, and it also depends on $n$ polarization tensors $\epsilon_{\mu\nu}=\epsilon_\mu\epsilon_\nu$. For example the four point amplitude has a schematic form,
\begin{equation}
\frac{(\epsilon\cdot p)^4(\epsilon\cdot\epsilon)^2}{s} + \frac{(\epsilon\cdot p)^4(\epsilon\cdot\epsilon)^2}{t} + \frac{(\epsilon\cdot p)^4(\epsilon\cdot\epsilon)^2}{u} + (\epsilon\cdot p)^2(\epsilon\cdot\epsilon)^6
\end{equation}
For each diagram we write an ansatz for the numerator $N_n^{(\Gamma)}$ with free parameters and impose the gauge invariance condition in $n-1$ external legs. As a result, we get an unique solution which is the graviton. Therefore, unitarity emerges from locality and gauge invariance in the same sense as in the Yang-Mills. The proof is very analogous to that case too using the soft limit and its uniqueness.

For gravity we can also make a stronger statement, that even locality emerges from gauge invariance. Assuming only the $n-3$ poles in the denominator, including multiple poles and with no reference to cubic graphs, we claim that the unique gauge-invariant is the amplitude.

We can also go back to the gluon case and consider now all possible $P^2$ poles, not just the ones with cyclic momenta, maintaining the non-trivial power-counting in the numerator, ie. $n-2$ momenta $p_j$, but now choosing $(n-3)$ of all possible cubic graph poles. Now imposing gauge invariance we conjecture $(n-2)!$ solutions corresponding to different cyclic orderings of Yang-Mills amplitudes modulo the relations following from the $U(1)$ decoupling and KK relations.

The uniqueness of gauge-invariants also gives a natural proof for the BCJ relation \cite{Bern:2008qj} between the Yang-Mills and gravity amplitudes. In particular, if we write the Yang-Mills amplitude in the BCJ form when for each cubic graph
the kinematical numerators satisfy $N_s+N_t=N_u$ if the color factor satisfy the same Jacobi identity, $c_s+c_t=c_u$.  Then the gravity amplitude is given by the simple replacement of the color factor by one more power of the kinematical numerator,
\begin{equation}
A_n^{(YM)} = \sum_\Gamma \frac{N_\Gamma c_\Gamma}{D_\Gamma}\quad\rightarrow\quad A_n^{(GR)} = \sum_\Gamma \frac{N_\Gamma^2}{D_\Gamma}\label{BCJ}
\end{equation}
The reason is very simple. Under a gauge variation, the $N_\Gamma$ change by some $\Delta_\Gamma$; the invariance of the full amplitude $\sum_\Gamma c_\Gamma \Delta_\Gamma/D_\Gamma = 0$ can then only be ensured by the Jacobi relations satisfied by $c_\Gamma$.
But if we now replace $c_\Gamma$ with some kinematical factor $N_\Gamma$ which satisfies the same identities, the gravity-gauge invariance check follows in exactly the same way as for YM. Thus the object with $c_{\Gamma} \to N_{\Gamma}$ is a gravitational gauge-invariant with $2(n-2)$ powers of momenta in the numerator; since this object is unique it gives the gravity amplitude.

\section{Gauge-Invariance $\to$ Soft Limits and Goldstone Theories}

We have seen that gauge and gravity amplitudes are much more special than garden-variety scalar theories. But of course famously there is also no good reason to have light scalars to begin with, unless they are goldstone bosons whose mass is appropriately protected by shift symmetries. Recent investigations revisiting some classic aspects of goldstone scattering amplitudes have revealed precisely what is special about these goldstone theories from a purely on-shell perspective. In the case of the non-linear sigma model, soft limit behavior in the form of the Adler zero \cite{adler} supplement unitarity and locality in certain cases to completely fix the tree-level S-matrix \cite{Cheung:2014dqa,Cheung:2015ota,Cheung:2016drk}. In particular, we can ask what is the minimally derivatively coupled theory which amplitudes have vanishing soft-limit, $A_n=0$ for $p_j\rightarrow0$. The answer appears to be non-linear sigma model (NLSM). If we demand the quadratic vanishing, $A_n={\cal O}(p^2)$ this uniquely specifies the Dirac-Born-Infeld (DBI) theory and $A_n={\cal O}(p^3)$ gives a special Galileon \cite{Cheung:2014dqa,Cachazo:2014xea}. The soft limit behavior was then used in the recursion relations to reconstruct the amplitudes in these theories, supplementing locality and unitarity.

In the spirit of our previous statements we can make the similar claims for these theories. Similar to Yang-Mills we can strip the flavor factor in the NLSM \cite{Kampf:2013vha} and consider cyclically ordered amplitudes $A_n$. Now the individual Feynman diagrams are quartic diagrams ${\cal Q}$, and we can write an expression for each of them
\begin{equation}
D_n^{({\cal Q})} = \frac{N_n^{({\cal Q})}(p_j)}{P_1^2P_2^2\dots P_{n/2-2}^2}\label{scal}
\end{equation}
Then the poles in (\ref{scal}) are cyclically ordered and the numerator is degree $n-2$ in momenta. Imposing the soft-limit vanishing then requires summing over all Feynman diagrams as only the amplitude has this property. Now we forget the Lagrangian and consider a general numerator,
\begin{equation}
N_n^{({\cal Q})}(p_j) = \sum_k \alpha_k\Delta_k \label{nums}
\end{equation}
where $\Delta_k$ is the product of $n/2-1$ terms of the form $s_{ij}=(p_i\cdot p_j)$. Note that if we allow one more $s_{ij}$ factor in the numerator then we could always write an expression which manifestly vanishes in the soft limit. For example, for the six point case one of the Feynman diagrams is
\begin{equation}
D = \frac{(s_{12}+s_{23})(s_{45}+s_{56})}{s_{123}}
\end{equation}
and it does not vanish in all soft limits, and no other numerator with two $s_{ij}$ does. If we replace the numerator by $s_{12}s_{34}s_{56}$ we would have manifestly each diagram vanishing.

Now we ask that the numerator $N_n^{({\cal Q})}$ is an arbitrary linear combination of products of $n/2-1$ factors $s_{ij}$ with free parameters. The statement is that imposing the soft limit vanishing in $n-1$ legs fixes all coefficients completely and there is an unique expression which is a $n$-pt tree-level amplitudes in NLSM. The proof for this statement uses double soft limit where two of the momenta go to zero. In that case the amplitude does not vanish but rather gives a finite expression, and in some sense it is an analogue of the Weinberg soft factor for the Yang-Mills and gravity. One can then prove the statement in a similar way to the soft limit argument for gluons and gravitons. The soft limit and locality then implies unitarity of goldstone amplitudes.

The stronger claim is that we do not have to consider quartic graphs, but rather take any expression with $n/2-2$ factors in the denominator (allowing double poles, and non-diagrammatic combinations of poles) and at most  $n/2-1$ terms $s_{ij}$ in the numerator. Then only imposing the soft limit again fixes the result uniquely, and we can see both locality and unitarity arising vanishing in the soft limit. We will give evidence for this  in the next section.

We can make analogous claims for the DBI and special Galileon. Now the power-counting of the numerator is $n-2$, resp. $3n/2-3$ factors $s_{ij}$ and $n/2-2$ poles in the numerator. We have to consider all quartic graphs with no ordering. Imposing the ${\cal O}(p^2)$, resp. ${\cal O}(p^3)$ vanishing in the soft limit of $n-1$ legs fixes the numerators uniquely to be the numerators of corresponding Feynman diagrams, and we get the amplitude as the only soft limit (with certain degree) vanishing object. The stronger statement again removes the requirement of single poles associated with quartic diagrams and we only consider the correct number $n/2-2$ poles $P^2$ with no restrictions.

\section{Evidence for the strong conjecture}

We have made two distinct claims: the first is that locality (in the form of the pole structures of cubic graphs), together with numerator power-counting, uniquely fixes the result when gauge-invariance/soft limits are imposed.

But we have also made a more striking conjecture, where we don't even impose locality, only ask that singularities are made of up to $(n-3)$ ``$P^2$" poles, without asking that these poles are associated with graphs at all.  And we demand the non-trivial number of momenta in the numerator which prohibits a trivial solution such as the powers of $(p_\mu\epsilon_\nu-p_\nu\epsilon_\mu)^{n}$ for gauge invariance of spin $s$, and the products of $\prod_j s_{j\,j+1}^\sigma$ for the soft limit ${\cal O}(p^\sigma)$. The claim is that the result is still unique; that locality and unitarity arise from (non-trivial) gauge-invariance/soft limits.

We do not currently have a proof of this conjecture, but if it is true, we suspect that the mechanism behind it should be the same for gluons, gravitons and goldstone theories. We will therefore confirm the conjecture for the case of the NLSM here; the way the graph structure emerges ``out of thin air" is already quite suggestive for what might be going on at general $n$.

For the 6pt NLSM amplitude there are only three poles $s_{123}$, $s_{234}$, $s_{345}$ which can appear in the denominator, and there is always just one of such factor. Therefore, locality here is directly imposed as we do not have any double poles or overlapping poles. The first non-trivial case to test our conjecture is then 8pt. The general ansatz is given by five different types of terms with two poles,
\begin{equation}
\widetilde{A}_8 = \frac{N_8^{(a)}}{s_{123}s_{456}}+\frac{N_8^{(b)}}{s_{123}s_{567}} +
\frac{N_8^{(c)}}{s_{123}s_{345}}+ \frac{N_8^{(d)}}{s_{123}s_{234}}+\frac{N_8^{(e)}}{s_{123}^2} \label{A8}
\end{equation}
Only the first two terms correspond to quartic graphs as the last three terms are not in the Feynman expansion as they violate locality. We will show that just soft limit vanishing forces $N_8^{(c)}=N_8^{(d)}=N_8^{(e)}=0$, or more precisely we can rewrite everything in terms of first two terms. Then we are left with the terms associated with quartic graphs only when the double soft limit argument can be applied to fix the answer uniquely to be an 8pt amplitude in NLSM.

The numerator is degree 6 in momenta, ie. degree 3 in invariants $s_{ij}$. It is convenient to use the cyclic basis,
\begin{equation}
s_{12},\dots,s_{81},s_{123},\dots,s_{812},s_{1234},\dots,s_{4567}.
\end{equation}
In the soft limit $p_8\rightarrow0$ these terms go to the 7pt cyclic basis made of $s_{12}$, $s_{123}$ and the cyclic images. Two of the terms $s_{78},s_{81}\rightarrow0$, the other nine terms $s_{23}$, $s_{34}$, $s_{45}$, $s_{56}$, $s_{234}$, $s_{345}$, $s_{456}$, $s_{2345}\rightarrow s_{671}$, $s_{3456}\rightarrow s_{712}$ stay the unique basis elements, while the remaining become degenerate (2-to-1 map).
\begin{align}
&s_{12},s_{812}\rightarrow s_{12};\,\, s_{1234},s_{567}\rightarrow s_{567};\,\, s_{4567},s_{123}\rightarrow s_{123}\nonumber \\
&\hspace{1cm} s_{678},s_{67}\rightarrow s_{67};\,\, s_{812},s_{12}\rightarrow s_{12}
\end{align}
Analogously for all other soft limits. It is very easy to show that $N_8^{(e)}=0$, or the corresponding term can be absorbed into first two terms in case we cancel one power of $s_{123}$. In the proof we critically use the relation between 7pt and 8pt basis of kinematical invariants. In the soft limit $\widetilde{A}_8=0$ and therefore the 7pt expression must vanish identically. Because the last term in (\ref{A8}) is the only term with that particular double pole in $s_{123}$ we apply different soft limits and demand that this term cancels or becomes degenerate with other terms.

For soft limits in momenta $p_2$, $p_5$, $p_6$, $p_7$ the term $s_{123}$ is a unique basis elements also in the 7pt basis, and there is no way how to cancel a double pole. Therefore, the numerator $N_8^{(e)}$ must simply vanish in all these four soft limits. For other four soft limits $s_{123}$ becomes degenerate with other kinematical invariants: with $s_{23}$ for $p_1\rightarrow0$, with $s_{12}$ for $p_3\rightarrow0$, with $s_{1234}$ for $p_4\rightarrow0$ and with $s_{4567}$ for $p_8\rightarrow0$. Therefore, either the numerator again vanishes or it is proportional to $s_{23}$ for $p_1\rightarrow0$, $s_{12}$ for $p_3\rightarrow0$ etc. It is easy to show that there is no such numerator $N_8^{(e)}$ which satisfies all these constraints. As a result, $N_8^{(e)}\sim s_{123}$ killing a double pole and being degenerate with first two terms.
The proofs for vanishing of $N_8^{(d)}$ and $N_8^{(c)}$ have the same flavor. 

It is likely that this sort of reasoning can be generalized to any $n$. Ultimately all statements about numerators $N_n^{(p)}$ are translated to properties of basis elements of kinematical invariants. The set of all cyclic invariants form a basis for any $n$ and they smoothly go to $n-1$ point basis in the soft limit. It seems plausible that some clever bookkeeping along the above lines can be done to prove the statement in general.

\section{Outlook}

There are a number avenues for further exploration suggested by this work.
One obvious question has to do with space-time dimensionality: all of our analysis find objects that would be gauge-invariant in {\it any} number of dimensions. But could there be functions that are only gauge-invariant in a specific dimension $d$?
In a specific space-time dimensionality $d$, there are further ``gram determinant" conditions that arise from the fact that any number $k>(d+1)$ momenta/vectors must be linearly dependent. Could it be that there are objects whose gauge-variation is proportional to gram determinant conditions in a specific number of dimensions, and so would be gauge-invariant in those dimensions but not otherwise? It is overwhelmingly likely that the answer to this question is ``no"--{\it all} non-trivial gauge-invariants are the ones that exist in all numbers of dimensions. This is certainly a fascinating feature of amplitudes for fundamental (parity-invariant) theories like YM and GR, and it would be nice to prove it directly along the lines of this note.

Resolving this issue about dimension-dependence would also settle a natural question posed by thinking about scattering amplitudes, the pursuit of which led directly to this work. Suppose we are given all the scattering amplitudes in some theory, these are  ``boundary observables in flat space"--they can be measured by experiments not in the interior of spacetime, but out at infinity. Given only this information, how could we discover the description of the physics in terms of local quantum evolution through the interior of the space-time? We can ask this question already at tree-level. We often say that the ``locality" and ``unitarity" of amplitudes is reflected in the location of their poles (locality) and the factorization of the poles on these poles (unitarity). But what we colloquially mean by these concepts is much more detailed that this--we would like to see that the amplitudes arise from local rules of particles moving and colliding at points in spacetime. Thus most prosaically, given the final amplitudes, we would like to know: how could we discover that they can be computed by the Feynman diagrams of a local theory?

As a trivial first step, we have to compare apples to apples. As we stressed in our introductory remarks the amplitudes are not Lorentz tensors, but Feynman amplitudes are. It is however trivial to associate on-shell amplitudes, written in terms of spinor-helicity variables, in terms of gauge-invariant Lorentz tensors. We can ``rationalize" any expression for amplitudes so that the poles are mandelstam invariants. Then e.g. an amplitude for a $-$ helicity  spin 1 particle would have weight two in it's $\lambda$ and is thus of the form $\lambda_\alpha \lambda_\beta T^{\alpha \beta}$ for some tensor $T^{\alpha \beta}$. But we can associate $\lambda_\alpha \lambda_\beta$ directly with a gauge invariant field strength; indeed defining $F^{\pm}_{\mu \nu} = F_{\mu \nu} \pm i \tilde{F}_{\mu \nu}$, we have that $F^{-}_{\alpha \dot{\alpha} \beta \dot{\beta}} = \lambda_\alpha \lambda_\beta \epsilon_{\dot{\alpha} \dot{\beta}}$, and similarly $F^{+}_{\alpha \dot{\alpha} \beta \dot{\beta}} = \epsilon_{\alpha \beta} \tilde{\lambda}_{\dot{\alpha}} \tilde{\lambda}_{\dot{\beta}}$. These expressions can be computed uniformly from $F_{\mu \nu} = p_{\mu} \epsilon_\nu - p_\nu \epsilon_{\mu}$, making the familiar choices for the helicity polarization vectors $\epsilon^{+}_{\alpha \dot{\alpha}} = \tilde \lambda_{\dot{\alpha}} \xi_{\alpha}/\langle\lambda \xi\rangle$ for any reference $\xi$, and similarly for $\epsilon^{-}_{\alpha \dot{\alpha}} = \lambda_\alpha \tilde{\xi}_{\dot{\alpha}}/[\tilde{\lambda} \tilde{\xi}]$.

In this way, any Lorentz-invariant expression of appropriate helicity weights can be associated with a gauge-invariant expression made out of field strengths. Consider for instance the 4 particle Parke-Taylor amplitude for $(1^- 2^- 3^+ 4^+)$, we can associate this with a gauge-invariant expression as
\begin{equation}
\frac{\langle12\rangle^2 [34]^2}{s t} \to \frac{(F^-_{1 \mu \nu} F_2^{- \mu \nu}) (F^+_{3 \alpha \beta} F_4^{+ \alpha \beta})}{s t}
\end{equation}
The right-hand side is non-vanishing only for this helicity configuration, so by summing over all helicities we construct a gauge-invariant expression that matches the amplitude on all helicity configurations.
It is amusing to carry out the exercise of constructing the Feynman amplitude from on-shell helicity amplitudes for the 4 gluon and 4 graviton amplitudes. Of course the individual helicity expressions explicitly involve $\epsilon_{\mu \nu \alpha \beta}$ and thus make sense only in four dimensions. But since the theory is parity invariant, after summing over all helicities all terms with an odd number of $\epsilon$'s cancel. Terms with an even number of $\epsilon$'s can be turned into expressions only involving the metric $\eta_{\mu \nu}$ using the fact that $\epsilon_{a b c d} \epsilon_{x y z w} = \left(\eta_{a x} \eta_{b y} \eta_{c z} \eta_{d w} \pm {\rm permutations} \right)$. This gives us
\begin{equation}
A_4 = \frac{(F_{\mu\nu} F^{\mu\nu})^2 - 2(F_{\mu\nu}F^{\nu\alpha}F_{\alpha\beta}F^{\beta\mu})}{s t}
\end{equation}
(where we expand $F_{\mu \nu} = F_{1 \mu \nu} + \cdots F_{4 \mu \nu}$ and we only keep terms linear in the polarization vectors). By construction matches the amplitude in four dimensions, but is a gauge-invariant expression in any number of dimensions.

Thus, starting from on-shell amplitudes, we can trivially construct the gauge-invariant Lorentz-tensor $M_{\mu_1 \cdots \mu_n}$ that matches all the helicity amplitudes making appropriate choices for the polarization vectors. We can now ask, how can we see that this object can be computed from local Feynman diagrams? Most naively, one might have expected that the critical properties of the amplitude--location of poles and factorization--would be needed in order to establish this fact. But we now see that much less is needed; even our weakest (and proven) statement about unique gauge-invariants requirements already shows that $M_{\mu_1 \cdots \mu_n}$ can be computed from Feynman diagrams. The reason is simply that Feynman diagrams give us a local gauge-invariant, and this object is unique, thus it must match the $M_{\mu_1 \cdots \mu_n}$ constructed from scattering amplitudes!

Note that however that complete proof of this fact requires us to show that there are no  gauge-invariants special to any particular dimension. [It is trivial to see that no ``gram determinant" conditions are possible for four-points, so the above expression is indeed valid in general $D$ dimensions, but this is no longer immediately true starting at five points].

Provided that the absence of dimension-specific invariants can be established, we have found a simple conceptual understanding of a fact that has resisted a transparent understanding for many years. There is an apparently straightforward proof that ``amplitudes that factorize properly" must match feynman diagrams, by using Cauchy's theorem and the BCFW deformation to show that that if functions have the same singularities they must be equal. However this famously needs a proof of an absence of poles at infinity on the Feynman diagram side, which can only be shown by a relatively indirect argument far afield from on-shell physics \cite{ArkaniHamed:2008yf}. 

The uniqueness of gauge invariance implies further properties of the S-matrices. In particular, it is trivial to show that it is impossible to have interactions of higher spin particles. The standard modern S-matrix argument relies on the factorization of the 4pt amplitude which is inconsistent in all three channels for $s>2$ \cite{Benincasa:2007xk,highspin}. In our story we do not use factorization, gauge invariance alone implies that any amplitude with cubic graphs needs a Weinberg soft factor, and that is impossible to construct for higher spins.

Our results also illuminate  why the CHY construction \cite{Cachazo:2013hca} of YM and gravity amplitudes must match the correct answer, without any detailed analysis of the poles and factorization structure. We simply observe the poles of the CHY formula are local, and the expressions are gauge invariant expression, with the correct units to match the correct numerator power-counting.

As we have seen the uniqueness of gauge-invariants gives a one-line proof of the passage from color-kinematics satisfying forms of Yang-Mills amplitudes to a gravity amplitude; it would be satisfying if the act of building gauge invariants naturally led to the color-kinematic structure for Yang-Mills to begin with.

Beyond these issues, the main open problem is to prove (or disprove) the strong conjecture about the emergence of the graph structure.Also, we have only looked at trees. Obviously the story at loop level will be much more interesting, and we can't expect uniqueness to follow simply from gauge invariance on external legs, since the particles propagating in the loops will also matter.

It is also natural to conjecture that maximal SUSY, together with degree $(n-3)$ poles, gives same result: diagrams emerge and the result is unique. We know that there is a tension between SUSY and locality/unitarity which is quite similar to the case of gauge invariance.

Finally, while the claims in this note are mathematically non-trivial and certainly have physical content, their ultimate physical significance is not clear. It is intriguing that locality and unitarity can be derived from the redundancy, inverting the usual logic leading to the need for gauge invariance. If this is more than a curiosity, it would be interesting to look for an abstract underlying system that gives rise to an effective description--either exactly or approximately--with a gauge redundancy, from which locality and unitarity emerge in the way we have seen here.

\smallskip

{\it Acknowledgment:}
We thank Freddy Cachazo, John Joseph Carrasco, David Kosower and Rutger Boels for stimulating discussions.
The work of NAH is supported in part by the US Department of Energy
under grant DOE DE-SC0009988.

\end{document}